\documentclass[conference]{IEEEtran}

\IEEEoverridecommandlockouts

\usepackage{url}
\usepackage{array}
\usepackage{cancel}
\usepackage{siunitx}
\usepackage{multirow}
\usepackage{graphicx}
\usepackage{textcomp}
\usepackage{multirow}
\usepackage{stfloats}
\usepackage{listings}
\usepackage{colortbl}
\usepackage{algorithm}
\usepackage[T1]{fontenc}
\usepackage[10pt]{moresize}
\usepackage[utf8]{inputenc}
\usepackage[noadjust]{cite}
\usepackage{mathtools, cuted}
\usepackage[noend]{algpseudocode}
\usepackage[table,x11names]{xcolor}
\usepackage{amsmath,bm,amssymb,amsfonts}

\def\BibTeX{{\rm B\kern-.05em{\sc i\kern-.025em b}\kern-.08em
		T\kern-.1667em\lower.7ex\hbox{E}\kern-.125emX}
}

\graphicspath{{./figs/}}

\definecolor{skyblue}{rgb}{0.53, 0.81, 0.92}
\definecolor{codegreen}{rgb}{0,0.6,0}
\definecolor{codegray}{rgb}{0.5,0.5,0.5}
\definecolor{codepurple}{rgb}{0.58,0,0.82}
\definecolor{backcolour}{rgb}{0.95,0.95,0.92}
\definecolor{LightCyan}{rgb}{0.88,1,1}
\definecolor{codeblue}{RGB}{49,49,255}
\definecolor{codeorange}{RGB}{255,143,102}
\definecolor{codewhite}{RGB}{255,255,255}
\definecolor{bittersweet}{rgb}{1.0, 0.44, 0.37}
\definecolor{columbiablue}{rgb}{0.61, 0.87, 1.0}
\definecolor{cornellred}{rgb}{0.7, 0.11, 0.11}

\lstdefinestyle{mystyle}{
	backgroundcolor=\color{backcolour}, 
	commentstyle=\color{codegreen},
	keywordstyle=\color{codepurple},
	numberstyle=\tiny\color{codegreen},
	stringstyle=\color{codepurple},
	identifierstyle=\color{cornellred},
	basicstyle=\ttfamily\footnotesize,
	breakatwhitespace=false,         
	breaklines=true,                 
	captionpos=t,                    
	keepspaces=true,                 
	numbersep=5pt,                  
	showspaces=false,                
	showstringspaces=false,
	showtabs=false,                  
	tabsize=2
}

\lstdefinestyle{plainstyle}{
	basicstyle=\ttfamily\footnotesize,
	keepspaces=true,                                
	numbersep=5pt,                  
	showspaces=false,                
	showstringspaces=false,
	showtabs=false,                  
	frame=tb,
	captionpos=b,
	tabsize=2,
	escapeinside={\%*}{*)}
}

\begin{document}
	\title{CAC~2.0: A Corrupt and Correct Logic Locking Technique Resilient to Structural Analysis Attacks \vspace*{-2mm}}
	
	\author{
		\IEEEauthorblockN{Levent~Aksoy\IEEEauthorrefmark{2}, Muhammad Yasin\IEEEauthorrefmark{3} and Samuel~Pagliarini\IEEEauthorrefmark{2}\IEEEauthorrefmark{4}}
		\IEEEauthorblockA{\IEEEauthorrefmark{2}Department of Computer Systems, Tallinn University of Technology, Tallinn, Estonia\\
			\IEEEauthorrefmark{3}Department of Computer and Software Engineering, National University of Sciences and Technology, Islamabad, Pakistan\\
			\IEEEauthorrefmark{4}Department of Electrical and Computer Engineering, Carnegie Mellon University, Pittsburgh, USA\\
			Email: levent.aksoy@taltech.ee, m.yasin@ceme.nust.edu.pk, pagliarini@cmu.edu}
			\vspace*{-6mm}
	}

	\maketitle
	
	\begin{abstract}
		Logic locking proposed to protect integrated circuits from serious hardware threats has been studied extensively over a decade. In these years, many efficient logic locking techniques have been proven to be broken. The \mbox{state-of-the-art} logic locking techniques, including the prominent corrupt and correct (CAC) technique, are resilient to satisfiability \mbox{(SAT)-based} and removal attacks, but vulnerable to structural analysis attacks. To overcome this drawback, this paper introduces an improved version of CAC, called CAC~2.0, which increases the search space of structural analysis attacks using obfuscation. To do so, CAC~2.0 locks the original circuit twice, one after another, on different nodes with different number of protected primary inputs using CAC, while hiding original protected primary inputs among decoy primary inputs. This paper also introduces an open source logic locking tool, called HIID, equipped with \mbox{well-known} techniques including CAC~2.0. Our experiments show that CAC~2.0 is resilient to existing SAT-based, removal, and structural analysis attacks. To achieve this, it increases the number of key inputs at most $4\times$ and the \mbox{gate-level} area between 30.2\% and 0.8\% on circuits with low and high complexity with respect to CAC.
	\end{abstract}
	
	\begin{IEEEkeywords}
		logic locking, obfuscation, SAT-based attack, removal attack, structural analysis
	\end{IEEEkeywords}
	
	\section{Introduction}

Due to the high cost of building a semiconductor foundry adopting the latest technology node, fabless design houses outsource the fabrication of their integrated circuits (ICs) to offshore foundries. They also integrate third-party intellectual properties (IPs) into their designs to reduce the time to market and implementation costs. However, such a globalized supply chain leads to serious hardware security threats, such as piracy, overproduction, and modification, which harm the semiconductor industry financially and undermine national security~\cite{dsb15}. To thwart such threats, many techniques, such as watermarking, digital rights management, metering, and logic locking~\cite{roy08}, have been introduced. Among those, logic locking has been a prominent solution to many security threats.

Logic locking inserts an additional logic with key inputs into the original design such that the locked design behaves as the original one only when the secret key is provided. Otherwise, it always generates a wrong output. In early years of logic locking, i.e., the pre-satisfiability (SAT) era, researchers explored different types of gates used for locking, such as {\sc xor}/{\sc xnor} gates in random logic locking (RLL)~\cite{roy08} and {\sc and}/{\sc or} gates~\cite{dupuis14}, considering the hardware complexity and output corruption. Then, the \mbox{SAT-based} attack~\cite{subramanyan15} broke all the logic locking techniques existing at that time. Its success lies in finding distinguishing input patterns (DIPs), which eliminate wrong key(s) in each iteration. In later years of logic locking, i.e., the post-SAT era, researchers explored different ways to generate locked circuits untractable for the SAT-based attacks by increasing the number of iterations in the SAT-based attack using a point function~\cite{yasin17} or by generating cyclic locked circuits that the SAT-based attack cannot directly handle~\cite{kaveh17}. However, removal, structural analysis, and certain \mbox{defense-specific} attacks~\cite{yasin20, sirone19, zhou17} overcome all these techniques as well. In this \mbox{cat-and-mouse} game, many logic locking techniques have been challenged and broken over the years.

Among the state-of-the-art logic locking techniques, the corrupt and correct (CAC) technique~\cite{kaveh19} is resilient to \mbox{SAT-based} and removal attacks. However, it is vulnerable to structural analysis attacks, which analyze the structure and function of subcircuit(s) of the locked design associated with key inputs. In this paper, we introduce the improved version of CAC, called CAC 2.0, where the original design is locked twice on different nodes with different number of protected primary inputs using CAC and decoy primary inputs are used to obfuscate original protected primary inputs. The aim of CAC~2.0 is to increase the search space of structural analysis attacks by means of obfuscation. It is implemented in a logic locking tool, called HIID, that can also implement other \mbox{well-known} logic locking techniques, e.g., \mbox{Anti-SAT}~\cite{xie19}, \mbox{SARLock}~\cite{yasin16}, \mbox{TTLock}~\cite{ttlock}, and compound logic locking with RLL~\cite{tan20}, at register transfer level (RTL). The main contributions of this paper are two-fold: (i)~a logic locking technique resilient to existing \mbox{SAT-based}, removal, and structural attacks; (ii)~an open source logic locking tool that includes prominent logic locking techniques. Experimental results show that CAC~2.0 is resilient to existing attacks and such resiliency is achieved by increasing the number of key inputs in the locked design up to 4$\times$ when compared to CAC under the same \mbox{SAT-based} attack resiliency. It is observed on circuits with low and high complexity that it increases the \mbox{gate-level} area of locked designs between 74.8\% and 2.2\% with respect to that of original designs.

The remainder of this paper is organized as follows: Section~\ref{sec:background} presents background concepts and gives the related work. CAC 2.0 is described along with HIID in Section~\ref{sec:tool}. Experimental results are presented in Section~\ref{sec:results} and finally, Section~\ref{sec:conclusion} concludes the paper.

	\section{Background}
\label{sec:background}

\subsection{Logic Locking}

\begin{figure*}[t]
	\centerline{\includegraphics[width=17.5cm]{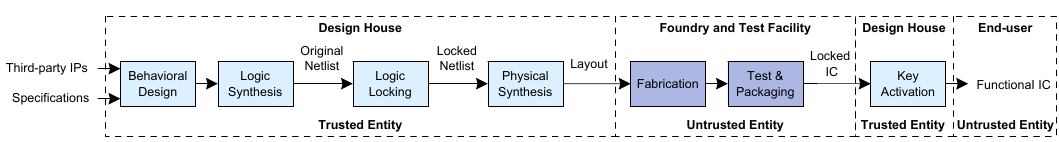}}
	\vspace*{-4mm}
	\caption{Conventional logic locking in the IC design flow.}
	\label{fig:iclock}
	\vspace*{-4mm}
\end{figure*}

In the IC design flow, logic locking can be applied at different stages, where the gate-level is the most common as shown in Fig.~\ref{fig:iclock}. In this scenario, the layout of the locked circuit is sent to the foundry without revealing the secret key. After the locked IC is fabricated and delivered to the design house, values of the secret key are stored in a \mbox{tamper-proof} memory before the functional IC is sent to the market. The adversary can be an individual at an untrusted foundry or an end-user, who has the ability to obtain the locked netlist by \mbox{reverse-engineering} the layout or functional IC, respectively. In the \mbox{oracle-less (OL)} threat model, it is assumed that the adversary has only the locked netlist. In the \mbox{oracle-guided (OG)} threat model, the adversary has also the functional IC, which can be used as an oracle to apply inputs and observe outputs.

\begin{figure}[t]
	\vspace*{-2mm}
	\centerline{\includegraphics[width=8.5cm]{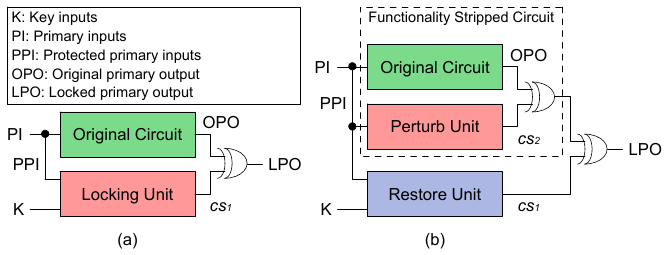}}
	\vspace*{-4mm}
	\caption{State-of-the-art logic locking techniques~\cite{aksoy24}: (a)~SFLT; (b)~DFLT.}
	\label{fig:sdflt}
	\vspace*{-6mm}
\end{figure}

\subsection{Defenses}
\label{subsec:defenses}

The state-of-the-art logic locking techniques resilient to the SAT-based attack are generally grouped in two classes: single and double flip locking techniques (SFLTs and DFLTs). They take advantage of the one-point function that evaluates to 1 at exactly one input pattern. Thus, they force the \mbox{SAT-based} attack to take an exponential number of iterations in terms of the number of key inputs since a DIP found by the \mbox{SAT-based} attack in an iteration eliminates only one wrong key~\cite{yasin17}. As shown in Fig.~\ref{fig:sdflt}(a), SFLTs use the critical signal $cs_1$, which corrupts the original circuit for wrong keys. For example, \mbox{Anti-SAT} utilizes complementary functions, which are generally composed of an {\sc and} gate tree, whose output is merged with the original circuit and \mbox{SARLock} adds a comparator and a masking circuit connected with the original netlist. \mbox{CAS-Lock~\cite{shakya19}} is based on the same concept of \mbox{Anti-SAT}, but uses a mix of {\sc and} and {\sc or} gates. As shown in Fig.~\ref{fig:sdflt}(b), DFLTs use the critical signal $cs_2$ to corrupt the original design for a specific input and use the critical signal $cs_1$ to correct this corruption. The TTLock, CAC, and stripped functionality logic locking (SFLL) techniques~\cite{yasin17,sengupta20} are notable DFLTs. Listing~\ref{lst:cac} presents the description of CAC at RTL, where \textit{PPI} and \textit{K} are the $n$-bit protected primary inputs and key inputs, respectively, \textit{OPO} and \textit{LPO} are the original and locked primary outputs, respectively, and \textit{SK} is the $n$-bit secret key. 

\begin{lstlisting}[mathescape=true, float=t, language=verilog, caption={CAC described at RTL.}, label=lst:cac]
	module CAC (PPI, K, OPO, LPO);
	
	input [n-1:0] PPI, K;
	input OPO;
	output LPO;
	
	reg corrupt_out, correct_out;
	wire [n-1:0] SK = 123...; //secret key
	
	//Corrupt unit
	always @(*) begin 
	  corrupt_out = OPO;
	  if (PPI == SK) begin
	    corrupt_out = !(corrupt_out);
	  end
	end 
	//Correct unit
	always @(*) begin 
	  correct_out = corrupt_out;
	  if (PPI == K || PPI == SK) begin
	    correct_out = !(correct_out);
	  end
	end 
	
	assign LPO = correct_out;
	
	endmodule
\end{lstlisting}

The main drawback of using the one-point function is that the locked design differs from the original one on only one input pattern, leading to a very small output corruption. Hence, Gen-Anti-SAT~\cite{zhou21} uses \mbox{non-complementary} functions, \mbox{SFLL-HD$^h$}~\cite{yasin17} considers protected input patterns under the given Hamming distance value $h$, and the \mbox{multi-flip} locking technique, called strong Anti-SAT~\cite{yuntao20}, flips multiple nodes to increase the output corruption. Note that there also exist compound logic locking techniques~\cite{tan20}, including one, which leads to a high output corruption, such as RLL, and the other, which is resilient to the SAT-based and removal attacks, such as SFLL-Flex~\cite{sengupta20}. 

The combination of logic locking and obfuscation has also been a common practice to protect specific IPs and increase the attack effort by misleading the adversary~\cite{zhou19,aksoy23}. The techniques of~\cite{yasin17,rcalut} hide the functionality of the restore unit of DFLTs depicted in Fig.~\ref{fig:sdflt}(b) in read-proof hardware. 

\subsection{Attacks}
\label{subsec:attacks}

Under the OL threat model, there exist attacks, which generally explore patterns in the structure of a locked netlist using statistical analysis~\cite{li19,zhang19,alaql21}. For example, the SCOPE attack~\cite{alaql21} is an unsupervised constant propagation technique, which analyzes each key bit of the locked design after it is assigned to logic 0 and 1 value for critical features, such as area, delay, and power dissipation, which can reveal its correct value. KRATT~\cite{aksoy24} determines the secret key of SFLTs by finding values of key inputs in the locking unit, which set its output $cs_1$ to a constant value for all possible values of protected primary inputs using quantified Boolean formula (QBF). It makes a strong guess on the secret key of DFLTs after it finds the mapping between protected primary inputs and key inputs in the restore unit, replaces these primary inputs with related key inputs in the functionality stripped circuit (FSC), and runs the SCOPE attack.

Removal attacks~\cite{yasin20} can find the single critical signal $cs_1$ in SFLTs as shown in Fig.~\ref{fig:sdflt}(a) and obtain the original circuit by removing the locking unit. Note that the perturb unit of DFLTs, and consequently, the original circuit, can be identified if FSC is synthesized as two separate blocks as shown in Fig.~\ref{fig:sdflt}(b). The security diagnostic tool Valkyrie~\cite{limaye22} determines the vulnerability of SFLTs and DFLTs and finds critical signals. 

Under the OG threat model, many variants of the \mbox{SAT-based} attack with different techniques and aims have been developed. For example, the technique of~\cite{shen17}, called Double DIP, eliminates at least 2 DIPs in a single iteration. The attack of~\cite{shamsi17}, called AppSAT, aims for approximate functional recovery. 

The structural analysis attack of~\cite{sirone19} targets \mbox{TTLock}. It initially finds the relation between protected primary inputs and key inputs from the restore unit. Then, it determines the secret key by finding logic cones with a set of gates in FSC including the same protected primary inputs, finding possible input values of these logic cones, and comparing the output of the locked design with that of the oracle under these input patterns. The sparse prime implicant (SPI) attack of~\cite{zhaokun21} targets DFLTs. It analyzes the prime implicant table (PIT) of the locked design, considering that the protected input pattern may and may not merge with the PIT of the original design. The attack of~\cite{patnaik22} targets both SFLTs and DFLTs. It generates test patterns satisfying the properties of a logic locking technique and tries those with the oracle to find the secret key. KRATT can also handle both SFLTs and DFLTs. For SFLTs, it uses the QBF technique mentioned under the OL threat model. For DFLTs, similar to other structural analysis attacks, it initially finds the mapping between protected primary inputs and key inputs. Then, it determines possible values of logic cones in FSC including only the protected primary inputs as primary inputs using a SAT formulation and tries them with the oracle.  

Note that CAC is resilient to removal attacks since FSC is generated by complementing an output of the original circuit on the protected input pattern determined by the secret key\footnote{Due to the selected protected primary inputs and protected input pattern, the original circuit may reside in FSC after logic synthesis. If such a rare case occurs, the original circuit can be locked with a different protected input pattern until FSC does not include the original circuit.}. It is resilient to SAT-based attacks since the locked circuit generates the same output under a wrong key as the original one for all input patterns except two determined by the wrong and secret keys. However, it is vulnerable to structural analysis attacks since they explore only one protected input pattern based on the mapping between protected primary inputs and key inputs, which can be easily extracted from the restore unit.

	\section{Proposed Logic Locking Technique and Tool}
\label{sec:tool}

In this section, we initially describe CAC~2.0, the improved version of CAC, and then, HIID, the logic locking tool. 

\subsection{CAC~2.0: Proposed Logic Locking Technique}
\label{subsec:cac}

As described in Section~\ref{subsec:attacks}, there are two main drawbacks of CAC against structural analysis attacks: (i)~there exists only one protected input pattern; (ii)~the mapping between protected primary inputs and key inputs can be extracted from the restore unit. To address these drawbacks, the technique, which locks the original circuit twice using CAC, is introduced in Section~\ref{subsec:doublecac}, while the technique, which obfuscates protected primary inputs, is described in Section~\ref{subsec:obfuscation}. 

\begin{figure}[t]
	\vspace*{-4mm}
	\centerline{\includegraphics[width=9.0cm]{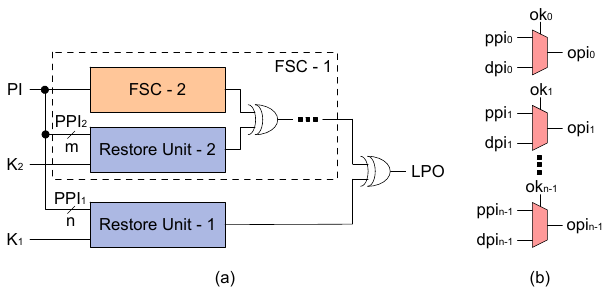}}
	\vspace*{-4mm}
	\caption{Techniques to overcome the vulnerability of CAC to structural analysis attacks: (a)~double CAC; (b)~obfuscation of protected primary inputs.}
	\label{fig:dcacobf}
	\vspace*{-6mm}
\end{figure}

\subsubsection{Double CAC}
\label{subsec:doublecac}

In order to increase the number of protected input patterns and force the adversary to find all of them to determine the secret key, the original circuit can be locked repeatedly using CAC on different nodes with different number of protected primary inputs. It is illustrated in Fig.~\ref{fig:dcacobf}(a) while applying CAC twice. Initially, an original primary output with at least $n$ primary inputs is locked by CAC using $n$ protected primary inputs $PPI_1$ and $n$ key inputs $K_1$ at RTL, the locked circuit is synthesized, and its \mbox{gate-level} netlist is obtained. Then, FSC of the locked circuit, \mbox{\textit{FSC-1}}, is identified, an output of a gate in \mbox{\textit{FSC-1}} is locked by CAC using $m$ protected primary inputs $PPI_2$ and $m$ key inputs $K_2$, where $m \leq n$, the locked circuit is synthesized, and its \mbox{gate-level} netlist is obtained. Note that the node in \textit{FSC-1} locked for the second time is different from the original primary output locked for the first time, but has the same $n$ protected primary inputs $PPI_1$ in its logic cone. The number of protected primary inputs in $PPI_2$ is suggested to be less than that of $PPI_1$, i.e., $m < n$, to increase output corruption. Note that the number of input patterns causing an output corruption under a wrong key is increased by $2^{n-m}$ with respect to CAC in this case.


\subsubsection{Obfuscation of Protected Primary Inputs}
\label{subsec:obfuscation}

In order to mislead the adversary while determining the mapping between protected primary inputs and key inputs in the restore unit and increase the number of logic gates to be explored in FSC, decoy primary inputs can be used to hide original protected primary inputs. To do so, a primary output with at least $2n$ primary inputs is selected to be locked, where $n$ of them are used as decoy primary inputs $dpi_{n-1}, \ldots, dpi_1, dpi_0$ while the other $n$ of them are used as protected primary inputs $ppi_{n-1}, \ldots, ppi_1, ppi_0$. To obfuscate each protected primary input with a decoy primary input, a MUX logic is described at RTL to select between these primary inputs based on an obfuscation key input as illustrated in Fig.~\ref{fig:dcacobf}(b). Note that such an obfuscation leads to $n$ additional MUXes and $n$ key inputs $ok_{n-1}, \ldots, ok_1, ok_0$, doubling the number of key inputs used in CAC. In the CAC module modified for obfuscation, an additional $n$-bit input \textit{OPI}, including the outputs of MUXes, $opi_{n-1}, \ldots, opi_1, opi_0$, is generated and it replaces \textit{PPI} in the \textit{if condition} of the correct unit in Listing~\ref{lst:cac}. Note that the function of an original primary output is still complemented on the protected input pattern in the corrupt unit in Listing~\ref{lst:cac}.

These techniques are combined in CAC~2.0, where the protected primary inputs $PPI_1$ and $PPI_2$ of the Double CAC technique are obfuscated accordingly. Thus, CAC~2.0 uses $k=2(m+n)$ key inputs, where $m+n$ key inputs are generated by the Double CAC technique while $m+n$ key inputs are generated by the obfuscation technique and the $k$-bit secret key is determined randomly. While $n$ key inputs are used to ensure resiliency against SAT-based attacks, $n+2m$ key inputs are used for obfuscation against structural analysis attacks.

CAC~2.0 increases the search space of structural analysis attacks due to two reasons. First, since the mapping between protected primary inputs and key inputs is not obvious to the attacker, all nodes in FSC, whose logic cone includes decoy and protected primary inputs as primary inputs, need to be considered. Second, since the locked circuit has two protected input patterns $PPI_1$ and $PPI_2$, they need to be explored together to find the secret key formed as the combination of $K_1$, $K_2$, and their obfuscation key inputs $OK_1$ and $OK_2$. The number of input patterns to be tried with the oracle is computed as the multiplication of all possible $PPI_1$ and $PPI_2$ patterns, leading to a maximum of $2^{2n+2m}$ input patterns.

\subsection{HIID: Proposed Logic Locking Tool}
\label{subsec:hiid}

Our logic locking tool, HIID, automates the process of locking an original circuit using the state-of-the-art techniques including CAC~2.0. It supports circuits in netlist bench format and Verilog. It requires the logic locking technique, the number of key inputs and protected input patterns, and other additional parameters as inputs. It describes logic locking techniques at RTL, which is one level higher than the gate-level in the IC design flow, since the description of logic to be locked and obfuscated gets easier and the logic synthesis of the locked design is more effective. It initially finds the logic cone of each primary output, computes its number of primary inputs, and determines the primary output to be locked according to the given number of key inputs. Then, it generates the locked design in Verilog by extending the original design with the logic locking module, its instantiation, and the declaration of key inputs and additional variables required to generate the locked design. The locked design in Verilog is synthesized using a logic synthesis tool and its gate-level netlist is generated. The locked design under the secret key is verified formally against the original design. Currently, HIID is equipped with \mbox{Anti-SAT}, SARLock, TTLock, CAC, CAC~2.0 as described in Section~\ref{subsec:cac}, and compound logic locking with RLL. It is developed in Perl, uses Cadence Genus for logic synthesis, Cadence Conformal for formal verification, the ABC tool~\cite{abc} to convert files in between bench and Verilog formats, and the SAT solver cryptominisat~\cite{cryptominisat} for equivalence checking. It is available at~\cite{hiid}.

	\section{Experimental Results}
\label{sec:results}

As an experiment set, we used five circuits with low and high complexity from ISCAS'85 and ITC'99 benchmarks. Table~\ref{tab:bench} shows their details, where \textit{inputs} and \textit{outputs} are the number of inputs and outputs, respectively, while \textit{area}, \textit{delay}, and \textit{power} denote respectively total area in $\mu m^2$, delay in the critical path in $ps$, and total power dissipation in $\mu W$ obtained after logic synthesis using a commercial 65\;nm gate library.

\begin{table}[t]
	\centering
	\caption{Details of the ISCAS'85 and ITC'99 circuits.}
	\vspace{-3mm}
	\begin{tabular}{|l|c|c||c|c|c|}
		\hline
		Circuit & inputs & outputs & area & delay & power\\ 
		\hline \hline
		c2670  & 157 & 64  & 544  & 1041 & 9882   \\
		c5315  & 178 & 123 & 1307 & 1352 & 26018  \\
		b14    & 277 & 299 & 4212 & 5688 & 85646  \\
		b15    & 485 & 519 & 6927 & 4146 & 75797  \\
		b20    & 522 & 512 & 9426 & 5705 & 225851 \\
		\hline
	\end{tabular}
	\label{tab:bench}
	\vspace{-6mm}
\end{table}

Table~\ref{tab:synth} presents the gate-level synthesis results of circuits locked by Anti-SAT, SARLock, TTLock, CAC, and CAC~2.0 using HIID, where $k$ is the total number of key inputs. In this experiment, we used the same number of key inputs for each logic locking technique to achieve the same SAT-based attack resiliency, i.e., a maximum of $2^{32}$ iterations. Thus, 32 key inputs are used for SARLock, TTLock, and CAC, 64 key inputs are used for Anti-SAT, and 96 key inputs are used for CAC~2.0, where $n$ and $m$ are 32 and 16, respectively. 

\begin{table*}[t]
	\centering
	\footnotesize
	\caption{Synthesis results of locked circuits.}
	\vspace{-3mm}
	\begin{tabular}{|l|c|c|c|c|c|c|c|c|c|c|c|c|c|c|c|}
		\hline
		\multirow{2}{*}{Circuit} & \multicolumn{3}{c|}{Anti-SAT ($k$=64)} & \multicolumn{3}{c|}{SARLock ($k$=32)} & \multicolumn{3}{c|}{TTLock ($k$=32)} & \multicolumn{3}{c|}{CAC ($k$=32)} & \multicolumn{3}{c|}{CAC~2.0 ($k$=96)} \\ 
		\cline{2-16}
		& area & delay & power & area & delay & power & area & delay & power & area & delay & power & area & delay & power \\
		\hline \hline		
		c2670  & 820  & 1201 & 14257  & 721  & 1169 & 13415  & 718  & 1082 & 12632  & 730  & 1096 & 12441  & 951  & 1456 & 15318  \\
		c5315  & 1562 & 1424 & 30084  & 1487 & 1320 & 28965  & 1490 & 1474 & 29591  & 1473 & 1389 & 28480  & 1679 & 1401 & 30940  \\
		b14    & 4462 & 5628 & 90692  & 4418 & 5524 & 86733  & 4419 & 5472 & 89290  & 4440 & 5458 & 90196  & 4555 & 5732 & 88593  \\
		b15    & 7167 & 4340 & 78991  & 7076 & 4389 & 77574  & 7090 & 4487 & 77619  & 7082 & 4368 & 77950  & 7241 & 4240 & 78417  \\
		b20    & 9645 & 5828 & 228224 & 9739 & 5156 & 230398 & 9562 & 5781 & 227785 & 9555 & 5462 & 227020 & 9635 & 6032 & 225334 \\
		\hline
	\end{tabular}
	\label{tab:synth}
	\vspace{-4mm}
\end{table*}

Observe from Tables~\ref{tab:bench} and~\ref{tab:synth} that the impact of a logic locking technique on the hardware complexity grows smaller as the complexity of the original circuit increases. For example, the area overhead of the logic locking techniques \mbox{Anti-SAT}, SARLock, TTLock, CAC, and CAC~2.0 on the \textit{c2670} (\textit{b20}) instance is obtained as 50.7\% (2.3\%), 32.5\% (3.3\%), 31.9\% (1.4\%), 34.1\% (1.3\%), and 74.8\% (2.2\%), respectively. The hardware complexity of circuits locked by \mbox{SARLock}, TTlock, and CAC is very close to each other and is generally better than those locked by \mbox{Anti-SAT} since more key inputs are required for Anti-SAT to achieve the same SAT-based attack resiliency with others. For the same reason, CAC~2.0 leads to locked circuits with the largest hardware complexity. The maximum and minimum area overhead of its locked circuits with respect to those locked by CAC are found as 30.2\% and 0.8\% obtained on the \textit{c2670} and \textit{b20} instances, respectively.

We used the OL attacks, SCOPE and KRATT, as well as the OG attacks, SAT-based, Double DIP, AppSAT, and KRATT, to break these locked circuits. Since Double DIP and AppSAT may return a wrong key in a single run, they were run multiple times with different settings. Note that Double DIP was not able to break any locked circuits and therefore, its results were omitted. We also ran the Valkyrie tool on circuits locked by CAC~2.0, but could not determine any critical signals due to the irregular structures of its locked circuits. Tables~\ref{tab:ol} and~\ref{tab:og} present results of OL and OG attacks, respectively. In Table~\ref{tab:ol}, \textit{cdk} and \textit{dk} are the number of correctly deciphered key inputs and deciphered key inputs, respectively. In Table~\ref{tab:og}, \textit{OoT} indicates that a solution could not be found due to the given time limit set to 2 days. In these tables, \textit{NoS} stands for no solution and the \mbox{run-time} of attacks are given in seconds. The attacks were run on a computing server including 32 Intel Xeon processing units at 3.9~GHz with 128~GB memory.

\begin{table*}[t]
	\centering
	\footnotesize
	\caption{Results of OL Attacks on Locked Circuits.}
	\vspace{-3mm}
	\begin{tabular}{|@{\hskip3pt}l@{\hskip3pt}|c@{\hskip3pt}|@{\hskip3pt}c@{\hskip3pt}|@{\hskip3pt}c@{\hskip3pt}|@{\hskip3pt}c@{\hskip3pt}|@{\hskip3pt}c@{\hskip3pt}|@{\hskip3pt}c@{\hskip3pt}|@{\hskip3pt}c@{\hskip3pt}|@{\hskip3pt}c@{\hskip3pt}|@{\hskip3pt}c@{\hskip3pt}|@{\hskip3pt}c@{\hskip3pt}|@{\hskip3pt}c@{\hskip3pt}|@{\hskip3pt}c@{\hskip3pt}|@{\hskip3pt}c@{\hskip3pt}|@{\hskip3pt}c@{\hskip3pt}|@{\hskip3pt}c|@{\hskip3pt}c@{\hskip3pt}|@{\hskip3pt}c@{\hskip3pt}|@{\hskip3pt}c@{\hskip3pt}|c@{\hskip3pt}|}
		\hline
		\multirow{3}{*}{Circuit} & \multicolumn{4}{c|@{\hskip3pt}}{Anti-SAT ($k$=64)} & \multicolumn{4}{c|@{\hskip3pt}}{SARLock ($k$=32)} & \multicolumn{4}{c|@{\hskip3pt}}{TTLock ($k$=32)} & \multicolumn{4}{c|@{\hskip3pt}}{CAC ($k$=32)} & \multicolumn{3}{c|}{CAC~2.0 ($k$=96)} \\ 
		\cline{2-20}
		& \multicolumn{2}{c|@{\hskip3pt}}{SCOPE} & \multicolumn{2}{c|@{\hskip3pt}}{KRATT} & \multicolumn{2}{c|@{\hskip3pt}}{SCOPE} & \multicolumn{2}{c|@{\hskip3pt}}{KRATT} & \multicolumn{2}{c|@{\hskip3pt}}{SCOPE} & \multicolumn{2}{c|@{\hskip3pt}}{KRATT} & \multicolumn{2}{c|@{\hskip3pt}}{SCOPE} & \multicolumn{2}{c|@{\hskip3pt}}{KRATT} & \multicolumn{2}{c|}{SCOPE} & \multirow{2}{*}{KRATT}\\
		\cline{2-19}
		& cdk/dk & CPU & cdk/dk & CPU & cdk/dk & CPU & cdk/dk & CPU & cdk/dk & CPU & cdk/dk & CPU & cdk/dk & CPU & cdk/dk & CPU & cdk/dk & CPU & \\
		\hline \hline		
		c2670  & 1/2 & 4.3  & 64/64 & 0.7  & 32/32 & 2.0 & 32/32 & 0.5  & 0/0 & 1.9 & 20/32 & 62.6 & 0/0 & 1.9 & 18/32 & 62.6 & 22/32 & 6.2  & \multicolumn{1}{c|}{NoS} \\
		c5315  & 2/3 & 4.5  & 64/64 & 2.0  & 32/32 & 2.5 & 32/32 & 1.5  & 0/0 & 2.3 & 18/32 & 62.9 & 0/0 & 2.4 & 19/32 & 62.6 & 19/21 & 6.7  & \multicolumn{1}{c|}{NoS} \\
		b14    & 2/3 & 7.4  & 64/64 & 16.4 & 32/32 & 3.9 & 32/32 & 3.9  & 0/0 & 3.7 & 20/31 & 64.5 & 0/0 & 4.5 & 19/32 & 64.1 & 15/28 & 11.0 & \multicolumn{1}{c|}{NoS} \\
		b15    & 2/4 & 9.7  & 64/64 & 32.8 & 32/32 & 5.1 & 32/32 & 29.0 & 0/0 & 5.7 & 16/32 & 66.1 & 0/0 & 5.4 & 17/32 & 66.3 & 15/24 & 14.9 & \multicolumn{1}{c|}{NoS} \\
		b20    & 4/8 & 13.1 & 64/64 & 57.6 & 32/32 & 6.8 & 32/32 & 51.1 & 3/3 & 6.9 & 16/30 & 67.5 & 1/1 & 6.7 & 20/32 & 67.6 & 10/20 & 19.6 & \multicolumn{1}{c|}{NoS} \\
		\hline
	\end{tabular}
	\label{tab:ol}
	\vspace{-4mm}
\end{table*}

\begin{table*}[t]
	\centering
	\footnotesize
	\caption{Results of OG Attacks on Locked Circuits.}
	\vspace{-3mm}
	\begin{tabular}{|@{\hskip3pt}l@{\hskip3pt}|c@{\hskip3pt}|@{\hskip3pt}c@{\hskip3pt}|@{\hskip3pt}c@{\hskip3pt}|@{\hskip3pt}c@{\hskip3pt}|@{\hskip3pt}c@{\hskip3pt}|@{\hskip3pt}c@{\hskip3pt}|@{\hskip3pt}c@{\hskip3pt}|@{\hskip3pt}c@{\hskip3pt}|@{\hskip3pt}c@{\hskip3pt}|@{\hskip3pt}c@{\hskip3pt}|@{\hskip3pt}c@{\hskip3pt}|@{\hskip3pt}c@{\hskip3pt}|@{\hskip3pt}c@{\hskip3pt}|@{\hskip3pt}c@{\hskip3pt}|@{\hskip3pt}c@{\hskip3pt}|}
		\hline
		\multirow{2}{*}{Circuit} & \multicolumn{3}{c|@{\hskip3pt}}{Anti-SAT ($k$=64)} & \multicolumn{3}{c|@{\hskip3pt}}{SARLock ($k$=32)} & \multicolumn{3}{c|@{\hskip3pt}}{TTLock ($k$=32)} & \multicolumn{3}{c|@{\hskip3pt}}{CAC ($k$=32)}  & \multicolumn{3}{c|}{CAC~2.0 ($k$=96)} \\ 
		\cline{2-16}
		& SAT & AppSAT & KRATT & SAT & AppSAT & KRATT & SAT & AppSAT & KRATT & SAT & AppSAT & KRATT & SAT & AppSAT & KRATT \\
		\hline \hline		
		c2670  & OoT & OoT & 1.5  & OoT & OoT   & 0.9  & OoT & OoT  & 71.0  & OoT & OoT & 67.7  & OoT & OoT & NoS \\
		c5315  & OoT & OoT & 3.2  & OoT & OoT   & 2.2  & OoT & OoT  & 69.5  & OoT & OoT & 70.8  & OoT & OoT & NoS \\
		b14    & OoT & OoT & 22.6 & OoT & OoT   & 17.2 & OoT & OoT  & 86.1  & OoT & OoT & 96.1  & OoT & OoT & NoS\\
		b15    & OoT & OoT & 25.8 & OoT & OoT   & 19.8 & OoT & OoT  & 140.4 & OoT & OoT & 156.1 & OoT & OoT & NoS\\
		b20    & OoT & OoT & 47.3 & OoT & 38567 & 34.4 & OoT & OoT  & 205.1 & OoT & OoT & 179.7 & OoT & OoT & NoS\\
		\hline
	\end{tabular}
	\label{tab:og}
	\vspace{-6mm}
\end{table*}

Observe from Table~\ref{tab:ol} that while SCOPE is able to decipher all key inputs of circuits locked by SARLock with 100\% accuracy, it can decipher a small number of key inputs of circuits locked by Anti-SAT, TTLock, and CAC. In CAC~2.0, it can decipher minimum (maximum) 20.8\% (33.3\%) of the number of key inputs with minimum (maximum) 50\% (90.4\%) of accuracy. On the other hand, KRATT can easily find the secret key of SFLTs, i.e., \mbox{Anti-SAT} and SARLock, and decipher almost all key inputs of DFLTs, i.e., TTLock and CAC, but without a success in finding the secret key. However, it cannot find the mapping between protected primary inputs and key inputs and thus, cannot find a solution in circuits locked by CAC~2.0.

Observe from Table~\ref{tab:og} that while AppSAT is successful only on the \textit{b20} instance locked by SARLock due to multiple runs, KRATT can break all the circuits locked by Anti-SAT, SARLock, TTLock, and CAC. However, none of these attacks can break the circuits locked by CAC~2.0. 

\begin{figure*}[t]
	\centering
	\vspace*{-3mm}
	\parbox{5.9cm}{\centerline{\includegraphics[width=6.5cm]{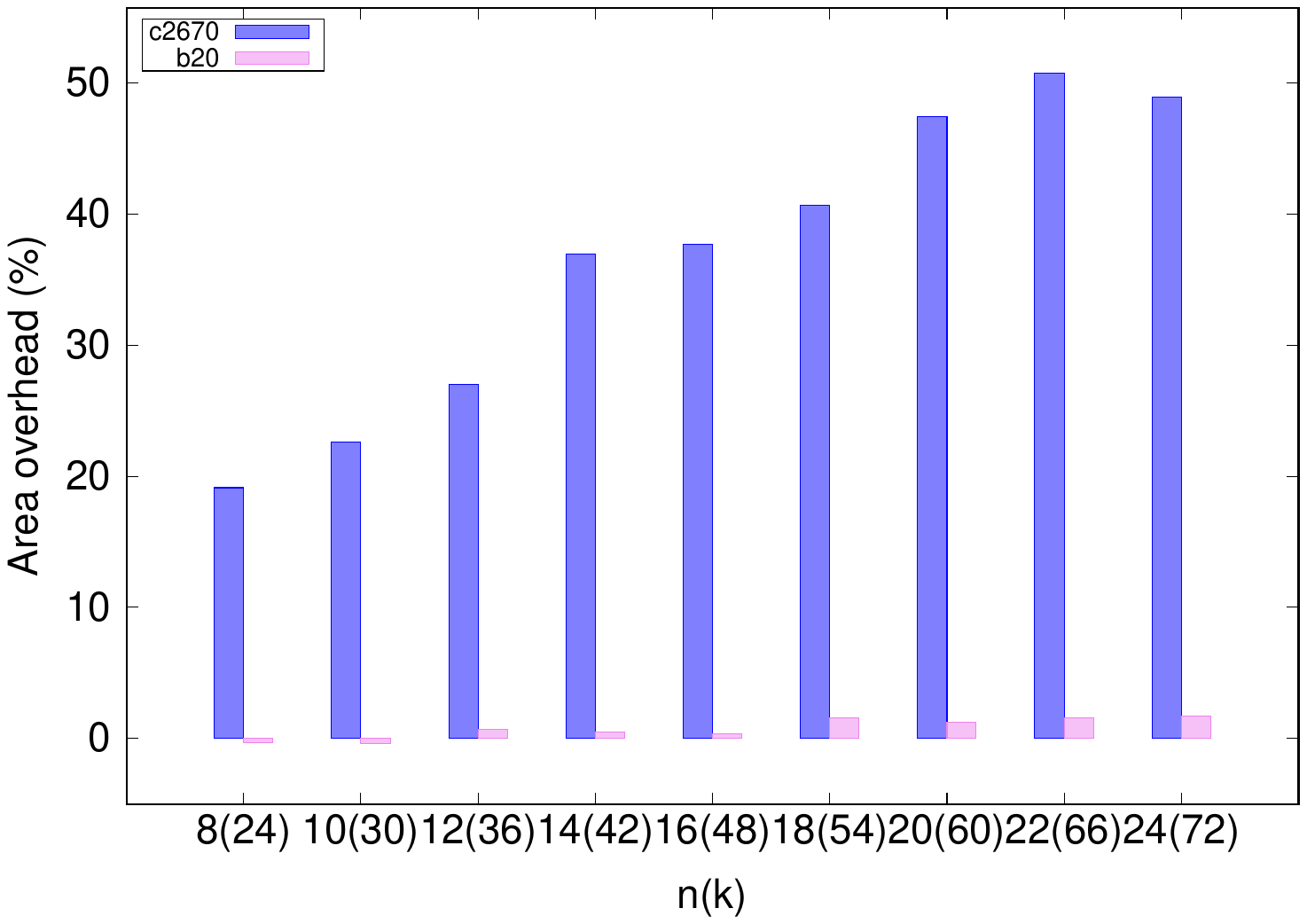}}}\
	\parbox{5.9cm}{\centerline{\includegraphics[width=6.5cm]{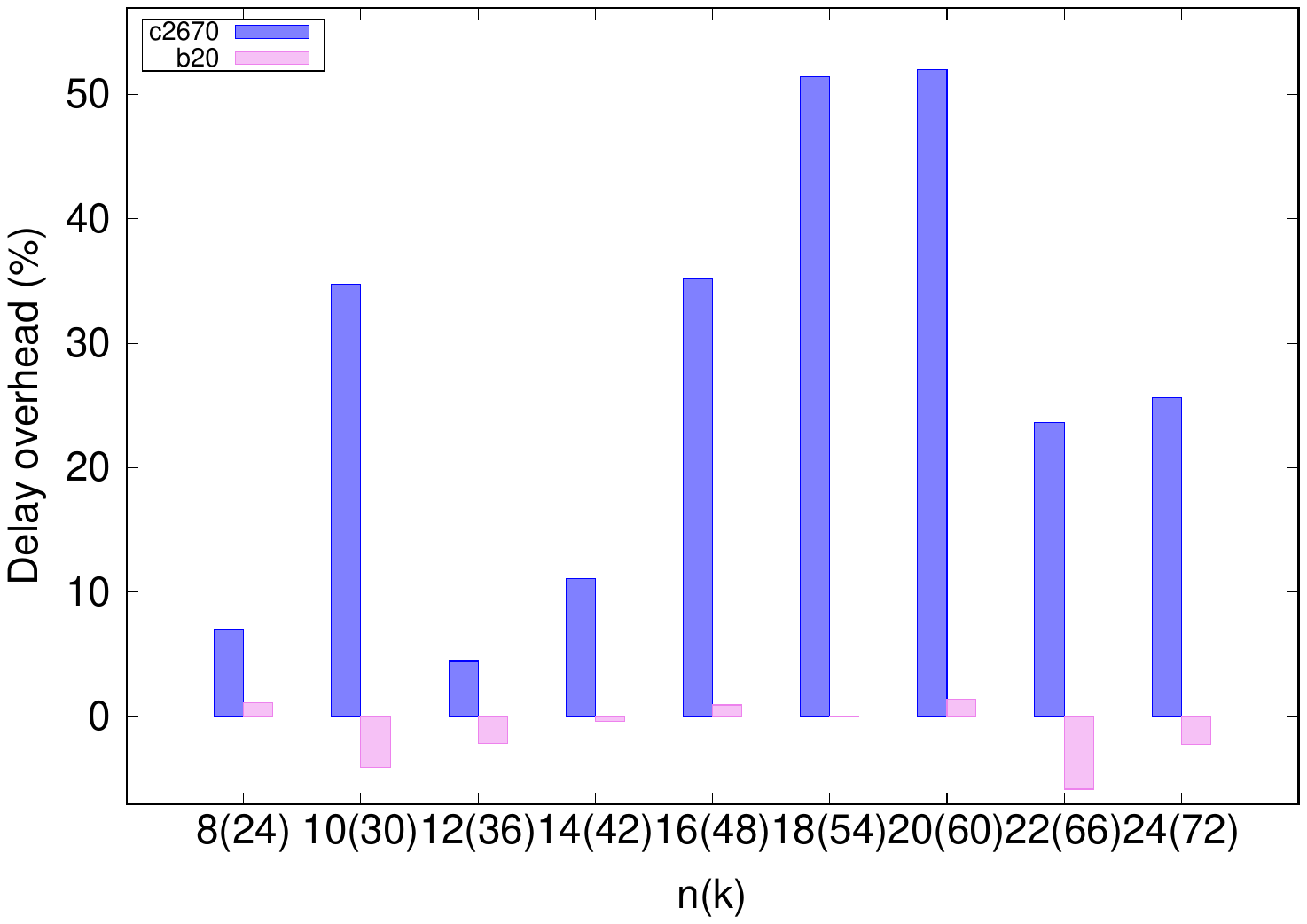}}}\
	\parbox{5.9cm}{\centerline{\includegraphics[width=6.5cm]{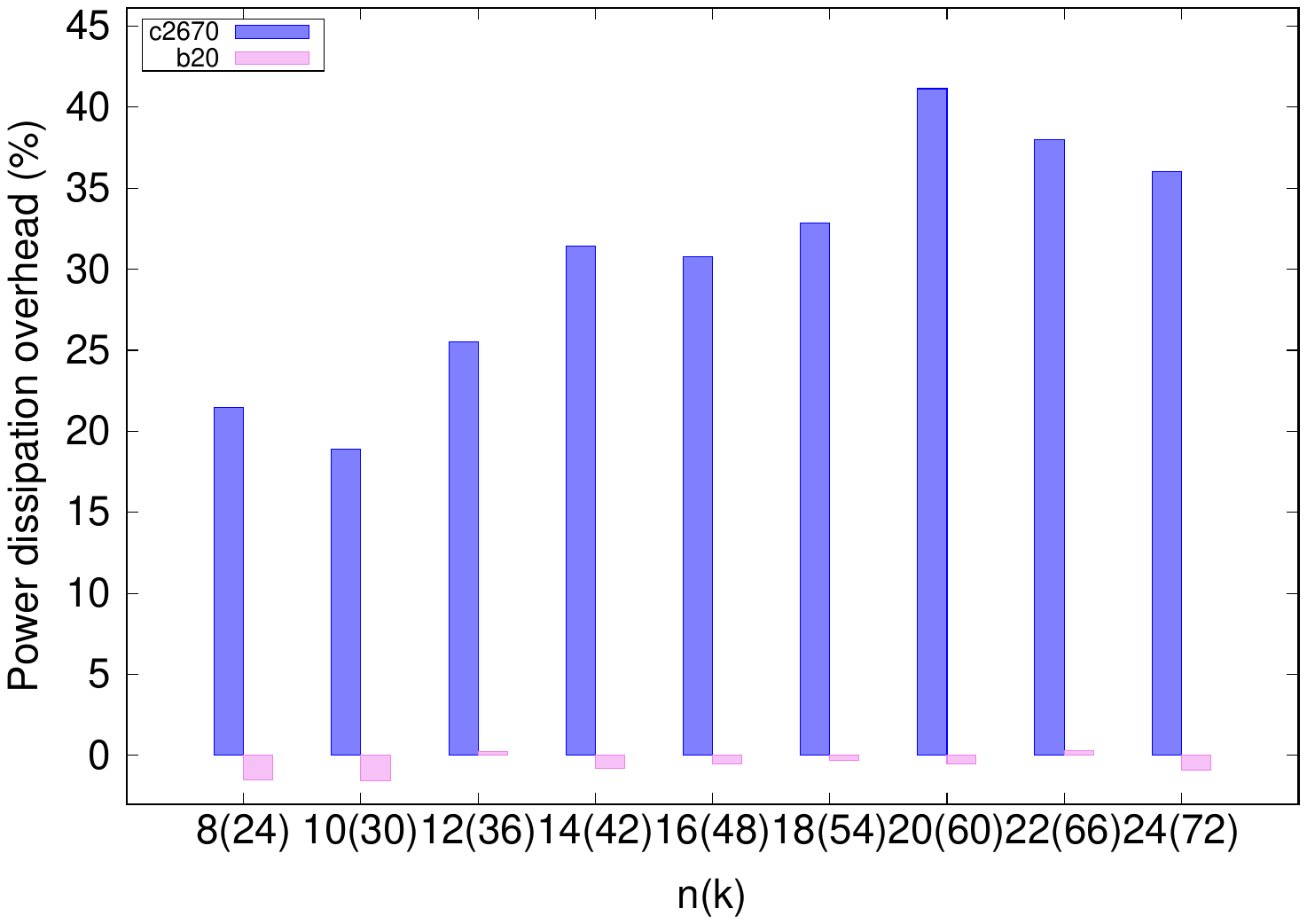}}}\
	
	\vspace*{-3mm}
	
	\parbox{5.9cm}{\centerline{\footnotesize (a)}}\
	\parbox{5.9cm}{\centerline{\footnotesize (b)}}\
	\parbox{5.9cm}{\centerline{\footnotesize (c)}}\
	\vspace*{-3mm}
	\caption{Impact of the number of key inputs on the hardware complexity in CAC~2.0: (a)~area; (b)~delay; (c)~power dissipation.}
	\label{fig:hc_nok}
	\vspace*{-6mm}
\end{figure*}

In order to find the impact of the number of key inputs on the overhead of the hardware complexity of a locked circuit with respect to the original one, the \textit{c2670} and \textit{b20} instances are locked using CAC~2.0 when $m = n/2$, where $n$ ranges between 8 and 24 in a step of 2, leading to a total number of key inputs $k=2(m+n)$ between 24 and 72 in a step of 6. These results are presented in Fig.~\ref{fig:hc_nok}. 

Observe from Fig.~\ref{fig:hc_nok} that although the overhead on hardware complexity increases as the number of key inputs increases on the \textit{c2670} instance, there are locked circuits with a small hardware complexity and a large number of key inputs. It is also observed on the \textit{b20} instance that the locked design may have better area, delay, and power dissipation values than the original one. This is simply due to the choice of the protected primary inputs and protected input patterns. 

\begin{figure}[t]
	\vspace*{-6mm}
	\centerline{\includegraphics[width=7.5cm]{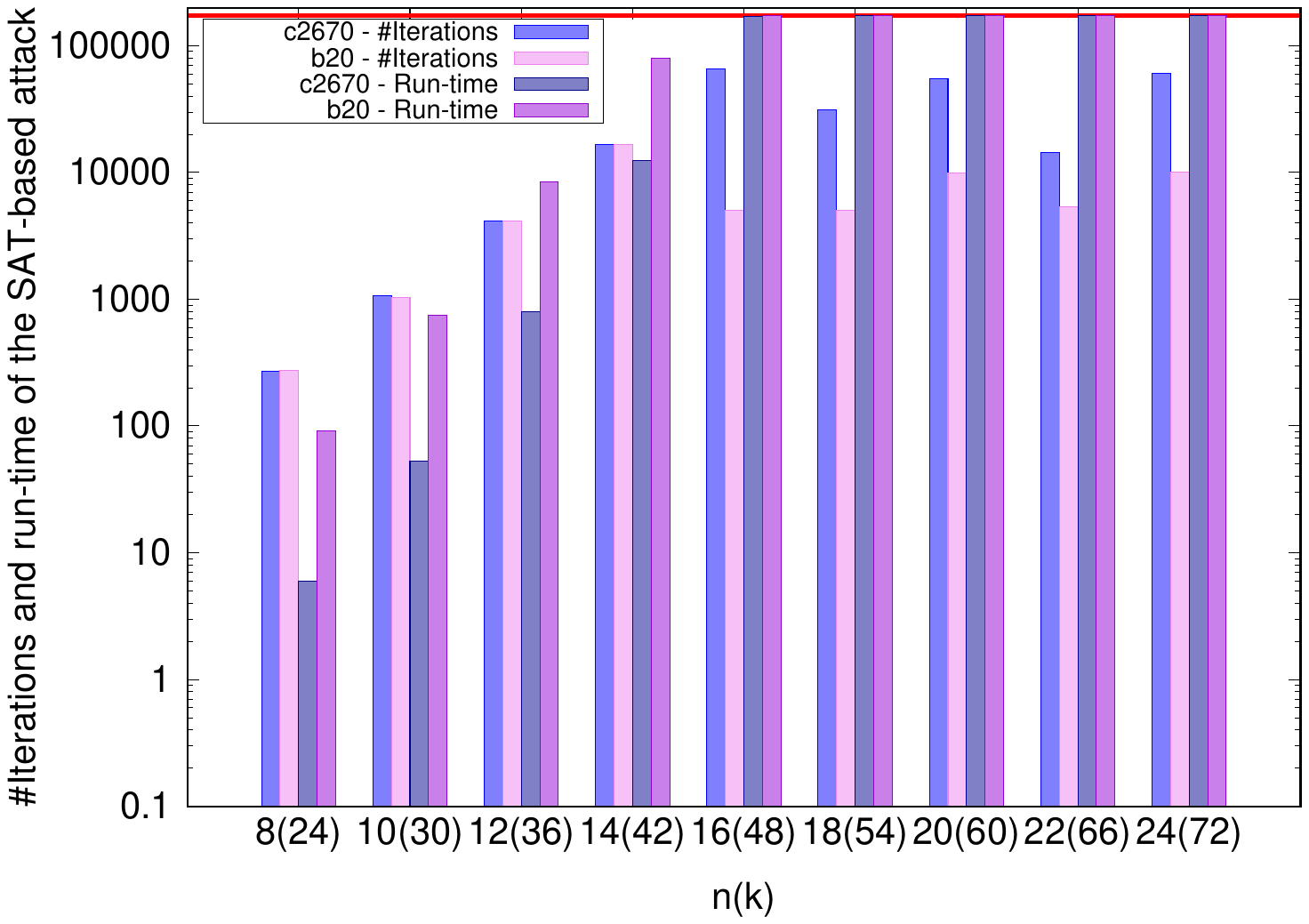}}
	\vspace*{-8mm}
	\caption{Impact of the number of key inputs on the number of iterations and run-time of the SAT-based attack in CAC~2.0.}
	\label{fig:sat_nok}
	\vspace*{-4mm}
\end{figure}

In order to find the impact of the number of key inputs on the SAT-based attack resiliency, the SAT-based attack was run on these locked circuits. Fig.~\ref{fig:sat_nok} presents its number of iterations and run-time on these locked circuits. The time limit was again 2 days as indicated by the red horizontal line in this figure.

Observe from Fig.~\ref{fig:sat_nok} that as the number of key inputs increases, the run-time of the SAT-based attack increases and the time limit is reached on the \textit{c2670} and \textit{b20} instances when $n$ ($k$) is greater than and equal to 18 (54) and 16 (48), respectively. The number of iterations increases exponentially with respect to the $n$ value until the given time limit is reached. In this case, the run-time of the SAT-based attack on the \textit{c2670} instance is smaller than that on the \textit{b20} instance simply because of less number of gates in the \textit{c2670} instance, and consequently, smaller SAT problem size.

	\section{Conclusions}
\label{sec:conclusion}

This paper introduced the multi-flip logic locking technique CAC~2.0, which is resilient to existing SAT-based, removal, and structural analysis attacks. It also presented the open source tool HIID, which is capable of locking an original circuit at RTL and is equipped with the \mbox{state-of-the-art} logic locking techniques including CAC~2.0. It is shown that CAC~2.0 achieves resiliency to all existing attacks by increasing the number of key inputs and gate-level area through obfuscation. 


	\section{Acknowledgment}

This work was supported by the EU through the European Social Fund in the context of the project “ICT programme”.

	\bibliographystyle{IEEEtran}
	\bibliography{lats24}

\begin{thebibliography}{10}
\providecommand{\url}[1]{#1}
\csname url@samestyle\endcsname
\providecommand{\newblock}{\relax}
\providecommand{\bibinfo}[2]{#2}
\providecommand{\BIBentrySTDinterwordspacing}{\spaceskip=0pt\relax}
\providecommand{\BIBentryALTinterwordstretchfactor}{4}
\providecommand{\BIBentryALTinterwordspacing}{\spaceskip=\fontdimen2\font plus
\BIBentryALTinterwordstretchfactor\fontdimen3\font minus
  \fontdimen4\font\relax}
\providecommand{\BIBforeignlanguage}[2]{{%
\expandafter\ifx\csname l@#1\endcsname\relax
\typeout{** WARNING: IEEEtran.bst: No hyphenation pattern has been}%
\typeout{** loaded for the language `#1'. Using the pattern for}%
\typeout{** the default language instead.}%
\else
\language=\csname l@#1\endcsname
\fi
#2}}
\providecommand{\BIBdecl}{\relax}
\BIBdecl

\bibitem{dsb15}
\BIBentryALTinterwordspacing
{Defence Science Board Task Force}. (2015, February) {On High Performance
  Microchip Supply Chain}. [Online]. Available:
  \url{https://dsb.cto.mil/reports/2000s/ADA435563.pdf}
\BIBentrySTDinterwordspacing

\bibitem{roy08}
J.~A. {Roy}, F.~{Koushanfar}, and I.~L. {Markov}, ``{EPIC: Ending Piracy of
  Integrated Circuits},'' in \emph{DATE}, 2008, pp. 1069--1074.

\bibitem{dupuis14}
S.~{Dupuis}, P.~{Ba}, G.~{Di Natale}, M.~{Flottes}, and B.~{Rouzeyre}, ``{A
  Novel Hardware Logic Encryption Technique for Thwarting Illegal
  Overproduction and Hardware Trojans},'' in \emph{IOLTS}, 2014, pp. 49--54.

\bibitem{subramanyan15}
P.~Subramanyan, S.~Ray, and S.~Malik, ``{Evaluating the Security of Logic
  Encryption Algorithms},'' in \emph{HOST}, 2015, pp. 137--143.

\bibitem{yasin17}
M.~Yasin, A.~Sengupta, M.~T. Nabeel, M.~Ashraf, J.~Rajendran, and O.~Sinanoglu,
  ``{Provably-Secure Logic Locking: From Theory To Practice},'' in \emph{ACM
  CCS}, 2017, pp. 1601--1618.

\bibitem{kaveh17}
K.~Shamsi, M.~Li, T.~Meade, Z.~Zhao, D.~Z. Pan, and Y.~Jin, ``{Cyclic
  Obfuscation for Creating SAT-Unresolvable Circuits},'' in \emph{GLVLSI},
  2017, p. 173–178.

\bibitem{yasin20}
M.~Yasin, B.~Mazumdar, O.~Sinanoglu, and J.~Rajendran, ``{Removal Attacks on
  Logic Locking and Camouflaging Techniques},'' \emph{IEEE TETC}, vol.~8,
  no.~2, pp. 517--532, 2020.

\bibitem{sirone19}
D.~Sirone and P.~Subramanyan, ``{Functional Analysis Attacks on Logic
  Locking},'' in \emph{DATE}, 2019, pp. 936--939.

\bibitem{zhou17}
H.~Zhou, R.~Jiang, and S.~Kong, ``{CycSAT: SAT-based Attack on Cyclic Logic
  Encryptions},'' in \emph{ICCAD}, 2017, pp. 49--56.

\bibitem{kaveh19}
K.~Shamsi, T.~Meade, M.~Li, D.~Z. Pan, and Y.~Jin, ``{On the Approximation
  Resiliency of Logic Locking and IC Camouflaging Schemes},'' \emph{IEEE TIFS},
  vol.~14, no.~2, pp. 347--359, 2019.

\bibitem{xie19}
Y.~Xie and A.~Srivastava, ``{Anti-SAT: Mitigating SAT Attack on Logic
  Locking},'' \emph{IEEE TCAD}, vol.~38, no.~2, pp. 199--207, 2019.

\bibitem{yasin16}
M.~{Yasin}, B.~{Mazumdar}, J.~{Rajendran}, and O.~{Sinanoglu}, ``{{SARLock}:
  {SAT} Attack Resistant Logic Locking},'' in \emph{HOST}, 2016, pp. 236--241.

\bibitem{ttlock}
M.~Yasin, A.~Sengupta, B.~C. Schafer, Y.~Makris, O.~Sinanoglu, and
  J.~Rajendran, ``{What to Lock? Functional and Parametric Locking},'' in
  \emph{GLSVLSI}, 2017, p. 351–356.

\bibitem{tan20}
\BIBentryALTinterwordspacing
B.~Tan \emph{et~al.}, ``{Benchmarking at the Frontier of Hardware Security:
  Lessons from Logic Locking},'' 2020. [Online]. Available:
  \url{https://arxiv.org/abs/2006.06806}
\BIBentrySTDinterwordspacing

\bibitem{aksoy24}
L.~{Aksoy}, M.~{Yasin}, and S.~{Pagliarini}, ``{KRATT: QBF-Assisted Removal and
  Structural Analysis Attack Against Logic Locking},'' in \emph{DATE}, 2024,
  accepted for publication.

\bibitem{shakya19}
B.~Shakya, X.~Xu, M.~Tehranipoor, and D.~Forte, ``{CAS-Lock: A
  Security-Corruptibility Trade-off Resilient Logic Locking Scheme},''
  \emph{IACR Transactions on Cryptographic Hardware and Embedded Systems}, vol.
  2020, no.~1, pp. 175--202, 2019.

\bibitem{sengupta20}
A.~{Sengupta}, M.~{Nabeel}, N.~{Limaye}, M.~{Ashraf}, and O.~{Sinanoglu},
  ``{Truly Stripping Functionality for Logic Locking: A Fault-Based
  Perspective},'' \emph{IEEE TCAD}, vol.~39, no.~12, pp. 4439--4452, 2020.

\bibitem{zhou21}
J.~Zhou and X.~Zhang, ``{Generalized SAT-Attack-Resistant Logic Locking},''
  \emph{IEEE TIFS}, vol.~16, pp. 2581--2592, 2021.

\bibitem{yuntao20}
Y.~Liu, M.~Zuzak, Y.~Xie, A.~Chakraborty, and A.~Srivastava, ``{Strong
  Anti-SAT: Secure and Effective Logic Locking},'' in \emph{ISQED}, 2020, pp.
  199--205.

\bibitem{zhou19}
H.~Zhou, A.~Rezaei, and Y.~Shen, ``{Resolving the Trilemma in Logic
  Encryption},'' in \emph{ICCAD}, 2019, pp. 1--8.

\bibitem{aksoy23}
L.~Aksoy, Q.-L. Nguyen, F.~Almeida, J.~Raik, M.-L. Flottes, S.~Dupuis, and
  S.~Pagliarini, ``{Hybrid Protection of Digital FIR Filters},'' \emph{IEEE
  TVLSI}, vol.~31, no.~6, pp. 812--825, 2023.

\bibitem{rcalut}
K.~Shamsi and Y.~Jin, ``{In Praise of Exact-Functional-Secrecy in Circuit
  Locking},'' \emph{IEEE TIFS}, vol.~16, no.~1, pp. 5225--5238, 2021.

\bibitem{li19}
L.~Li and A.~Orailoglu, ``{Piercing Logic Locking Keys through Redundancy
  Identification},'' in \emph{DATE}, 2019, pp. 540--545.

\bibitem{zhang19}
Y.~Zhang, P.~Cui, Z.~Zhou, and U.~Guin, ``{TGA: An Oracle-Less and
  Topology-Guided Attack on Logic Locking},'' in \emph{ASHES}, 2019, p.
  75–83.

\bibitem{alaql21}
A.~Alaql, M.~M. Rahman, and S.~Bhunia, ``{SCOPE: Synthesis-Based Constant
  Propagation Attack on Logic Locking},'' \emph{IEEE TVLSI}, vol.~29, no.~8,
  pp. 1529--1542, 2021.

\bibitem{limaye22}
N.~Limaye, S.~Patnaik, and O.~Sinanoglu, ``{Valkyrie: Vulnerability Assessment
  Tool and Attack for Provably-Secure Logic Locking Techniques},'' \emph{IEEE
  TIFS}, vol.~17, pp. 744--759, 2022.

\bibitem{shen17}
Y.~Shen and H.~Zhou, ``{Double DIP: Re-Evaluating Security of Logic Encryption
  Algorithms},'' in \emph{GLSVLSI}, 2017, pp. 179--184.

\bibitem{shamsi17}
K.~Shamsi, M.~Li, T.~Meade, Z.~Zhao, D.~Z. Pan, and Y.~Jin, ``{AppSAT:
  Approximately Deobfuscating Integrated Circuits},'' in \emph{HOST}, 2017, pp.
  95--100.

\bibitem{zhaokun21}
Z.~Han, M.~Yasin, and J.~J. Rajendran, ``{Does Logic Locking Work with {EDA}
  Tools?}'' in \emph{USENIX Security Symposium}, 2021, pp. 1055--1072.

\bibitem{patnaik22}
S.~Patnaik, N.~Limaye, and O.~Sinanoglu, ``{Hide and Seek: Seeking the
  (Un)-Hidden Key in Provably-Secure Logic Locking Techniques},'' \emph{IEEE
  TIFS}, vol.~17, pp. 3290--3305, 2022.

\bibitem{abc}
\BIBentryALTinterwordspacing
{A. Mishchenko}. {ABC: System for Sequential Logic Synthesis and Formal
  Verification}. [Online]. Available: \url{https://github.com/berkeley-abc/abc}
\BIBentrySTDinterwordspacing

\bibitem{cryptominisat}
\BIBentryALTinterwordspacing
{M. Soos}. {Cryptominisat SAT Solver}. [Online]. Available:
  \url{https://github.com/msoos/cryptominisat}
\BIBentrySTDinterwordspacing

\bibitem{hiid}
\BIBentryALTinterwordspacing
{L. Aksoy}. {HIID: A Logic Locking Tool}. [Online]. Available:
  \url{https://github.com/leventaksoy/hiid}
\BIBentrySTDinterwordspacing

\end{thebibliography}
\end{document}